\documentclass[a4paper]{jpconf}

\usepackage{graphicx}
\usepackage{amssymb}

\newcommand{\J}{JEM-EUSO}
\newcommand{\A}{PAO}
\newcommand\EeV{\rm{EeV}}

\newcommand{\beq}[1]{\begin{equation}\label{#1}}
\newcommand{\eeq}{\end{equation}}
\newcommand{\bea}[1]{\begin{eqnarray}\label{#1}}
\newcommand{\eea}{\end{eqnarray}}
\newcommand{\ba}{\begin{array}}
\newcommand{\ea}{\end{array}}

\newcommand{\rf}[1]{(\ref{#1})}
\newcommand{\rarr}{\rightarrow}
\newcommand{\be}{\begin{equation}}
\newcommand{\ee}{\end{equation}}

\begin{document}

\title{Sensitivity of orbiting JEM-EUSO to large-scale cosmic-ray anisotropies}

\author{
Peter B. Denton$^{1}$,
Luis A. Anchordoqui$^{2}$,
Andreas A. Berlind$^{1}$,
Matthew Richardson$^{1}$,
and 
Thomas J. Weiler$^{1}$,
for the JEM-EUSO Collaboration.
}

\address{
$^1$ Department of Physics \& Astronomy, Vanderbilt University, Nashville, TN, 37235 USA\\
$^2$ Physics Department, University of Wisconsin-Milwaukee, Milwaukee,  WI, 53201 USA
}

\def\myead#1{\vspace*{5pt}\address{Speaker's E-mail: \mailto{#1}}}
\myead{peterbd1@gmail.com}

\begin{abstract}
The two main advantages of space-based observation of extreme-energy ($\gtrsim 10^{19}$~eV) 
cosmic-rays (EECRs) over ground-based observatories 
are the increased field of view, and the all-sky coverage with nearly uniform systematics 
of an orbiting observatory.  The former guarantees increased statistics,
whereas the latter enables a partitioning of the sky into spherical harmonics.  
We have begun an investigation, using the spherical harmonic technique, 
of the reach of \J\ into potential anisotropies in the extreme-energy cosmic-ray sky-map.  
The technique is explained here, and simulations are presented.
The discovery of anisotropies would help to identify the long-sought origin of EECRs.
\end{abstract}

\section{Introduction}

The Extreme Universe Space Observatory (EUSO) is a consortium of 250 Ph.D. researchers from 27 institutions, 
spanning 14 countries.
It is a down-looking telescope optimized for near-ultraviolet fluorescence produced by extended air showers
in the atmosphere of the Earth.
EUSO is proposed to occupy the Japanese Experiment Module (JEM) on the International Space Station (ISS),
and collect up to 1000 cosmic ray (CR) events at and above 55 \EeV\ ($1\,\EeV=10^{18}$~eV) 
over a $5$~year lifetime, far surpassing the reach of any ground-based project.  

\J\ brings two new, major advantages to the search for the origins of EECRs.
One advantage is the large field of view (FOV), attainable only with a space-based observatory.
With a $60^\circ$ opening angle for the telescope, the down-pointing (``nadir'') FOV is 
\beq{nadirFOV}
\pi(h_{\rm ISS}\tan(30^\circ))^2\approx h_{\rm ISS}^2\approx 150,000\,{\rm km}^2\,.
\eeq
\A\ has a FOV of 3,000 km$^2$.
Thus the \J\ FOV, given in Eq.~(\ref{nadirFOV}), is 50 times larger for instantaneous measurements
(e.g., for observing transient sources).
Multiplying the \J\ event rate by an expected 18\% duty cycle, we arrive at a time-averaged
nine-fold increase in acceptance  for \J\ compared to \A, at energies where the \J\ efficiency
has peaked (at and above $\sim 100$ \EeV). 
Tilting the telescope turns the circular FOV given in Eq.~\rf{nadirFOV}
into a larger elliptical FOV.  The price paid for ``tilt mode'' is an increase
in the threshold energy of the experiment.

The second advantage is the coverage of the full sky ($4\pi$~steradians)
with nearly constant systematic errors on the energy and angle resolution, 
again attainable only with a space-based observatory.
(Combined data from ground-based observatories in the Northern and Southern hemispheres may offer full-sky coverage,
but not uniformity of systematics.) 
This talk pursues all-sky studies of possible spatial anisotropies.
The reach benefits from the $4\pi$~sky coverage, 
but also from the increased statistics resulting from the greater FOV.
A longer study will soon be completed and published~\cite{DABRW}.

In addition to the two advantages of space-based observation just listed, 
a third feature provided by a space-based mission may turn out to be significant.  
It is the increased acceptance for Earth-skimming neutrinos when the skimming 
chord transits ocean rather than land.
On this latter topic, just one study has been published~\cite{PalomaresRuiz:2005xw}.
The study concludes that an order of magnitude larger acceptance
results for Earth-skimming events transiting ocean compared to transiting land.
Ground-based observatories will not realize this benefit, 
since they  cannot view ocean chords.

\section{All-sky coverage and anisotropy}

As emphasized by Sommers over a dozen years ago~\cite{Sommers:2000us}, 
an all-sky survey offers a rigorous expansion in spherical harmonics,
of the normalized spatial event distribution
$I(\Omega)$, where $\Omega$ denotes the solid angle parameterized by
the pair of latitude ($\theta$) and longitude ($\phi$) angles:
\beq{SphHarm}
I(\Omega)\equiv \frac{N(\Omega)}{\int d\Omega\, N(\Omega)} =\sum_{\ell=0}^\infty \, \sum_{|m| \le l} a_{\ell m}\,Y_{\ell m}(\Omega)\,,
\eeq
i.e., the set $\{ Y_{\ell m} \}$ is complete. 
Averaging the $a^2_{\ell m}$ over the $(2\ell+1)$ values of $m$ defines the
rotationally-invariant ``power spectrum'' in the 
single variable $\ell$, 
$C(\ell)= \frac{1}{2\ell+1}\sum_{|m|\le\ell}\,a^2_{\ell m}$.
The set $\{ Y_{\ell m} \}$ is also orthogonal, obeying
\beq{ortho}
\int d\Omega \; Y_{{\ell_1} {m_1}}(\Omega) \, Y_{{\ell_2} {m_2}}(\Omega) =  
	\delta_{{\ell_1}{\ell_2}}\,\delta_{{m_1}{m_2}}\,.
\eeq
We are interested in the real valued $Y_{\ell m}$'s, defined as 
$P^\ell_m(x) (\sqrt{2}\cos(m\phi))$ for positive $m$, 
$P^\ell_{|m|}(x) (\sqrt{2}\sin(|m|\phi))$ for negative $m$, 
and $P_\ell(x)$ for $m=0$.
Here, $P^\ell_m$ is the associated Legendre polynomial, 
$P_\ell=P^\ell_{m=0}$ is the regular Legendre polynomial,
and $x\equiv \cos\theta$.

The lowest multipole is the $\ell=0$ monopole, 
equal to the average all-sky flux.
The higher multipoles ($\ell\ge 1$) and their amplitudes $a_{\ell m}$ correspond to anisotropies.
Guaranteed by the orthogonality of the $Y_{\ell m}$'s,
the higher multipoles when integrated over the whole sky equate to zero.

\begin{figure}[t]
\centering
\includegraphics[width=0.2\textwidth]{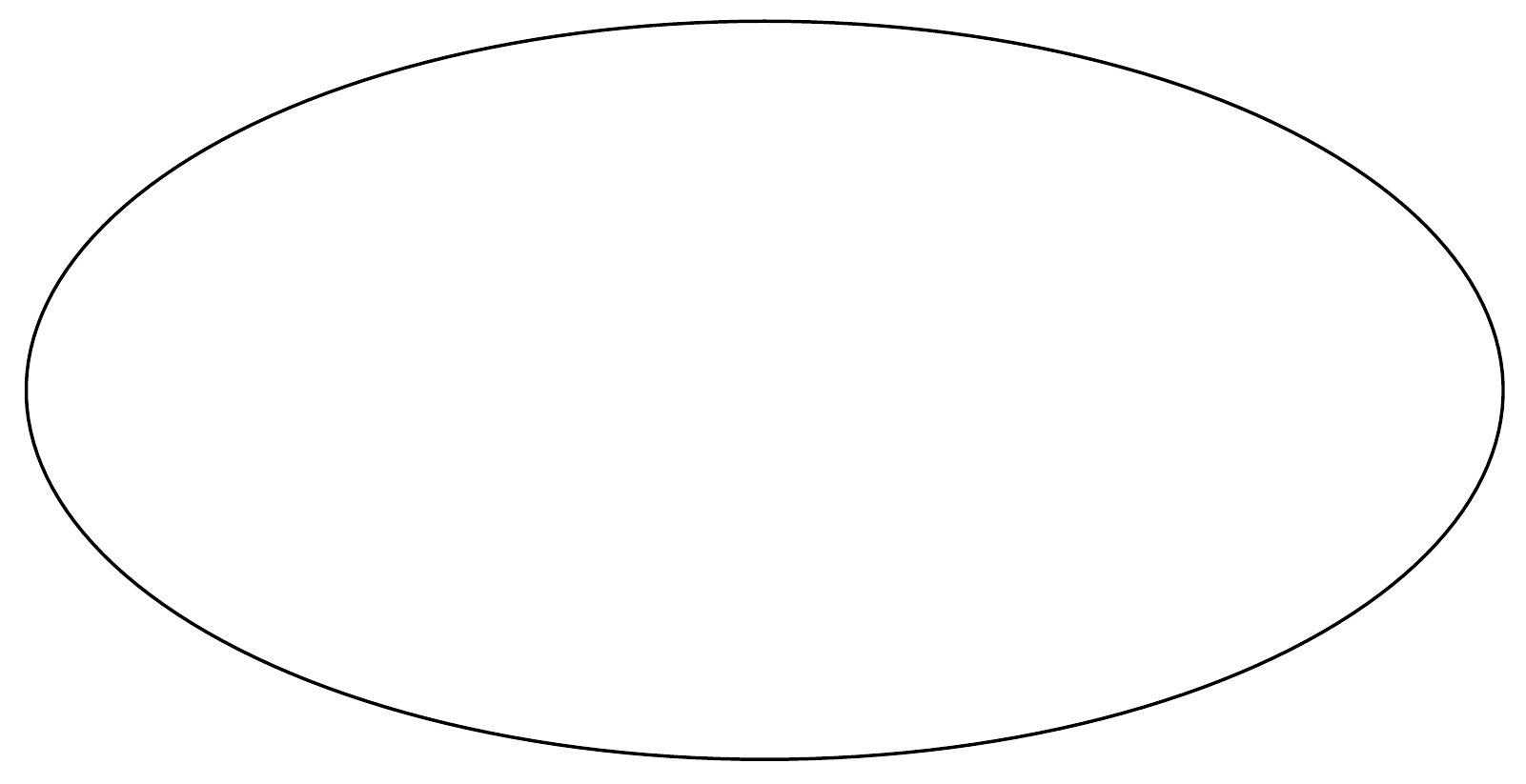}
\includegraphics[width=0.2\textwidth]{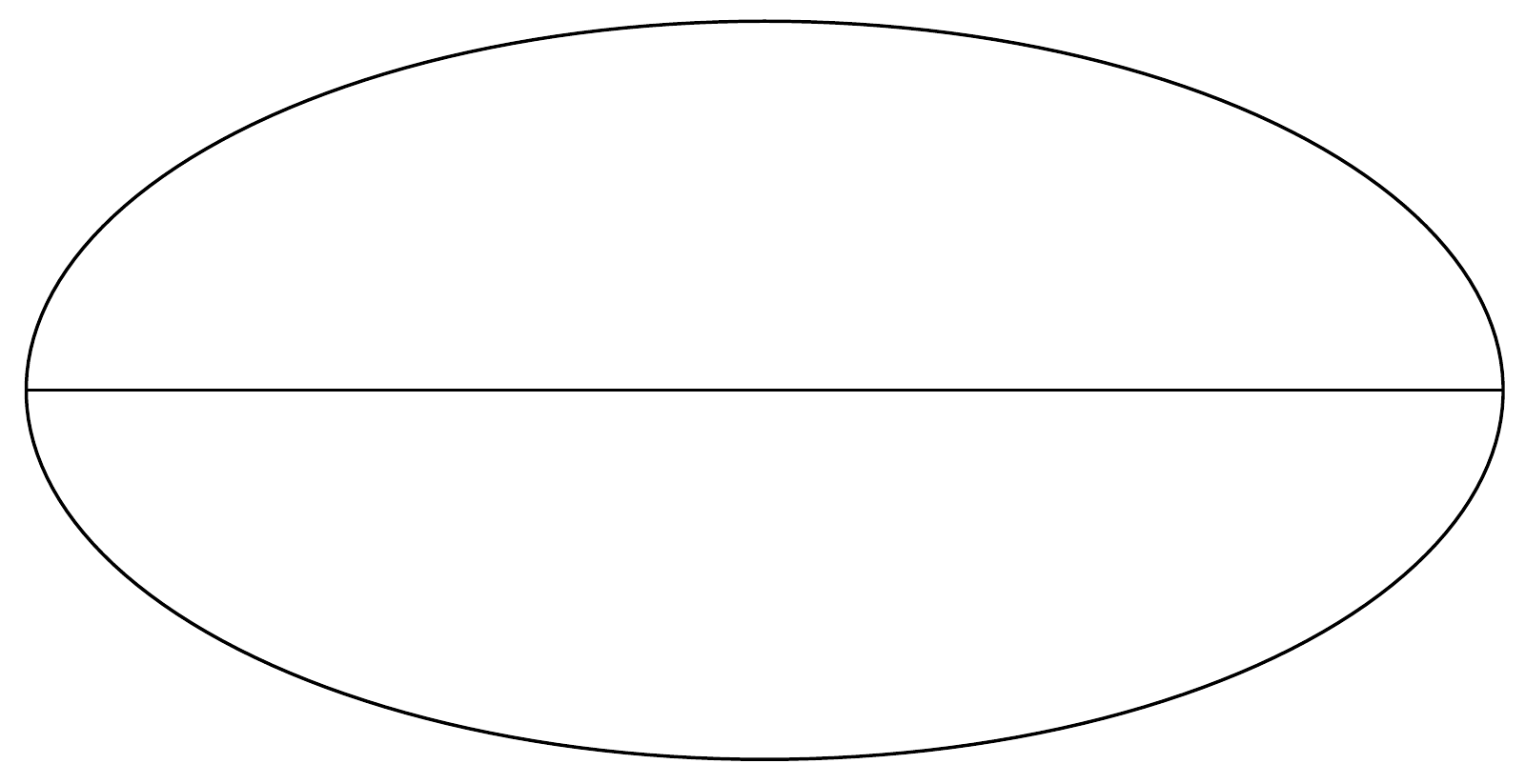}
\includegraphics[width=0.2\textwidth]{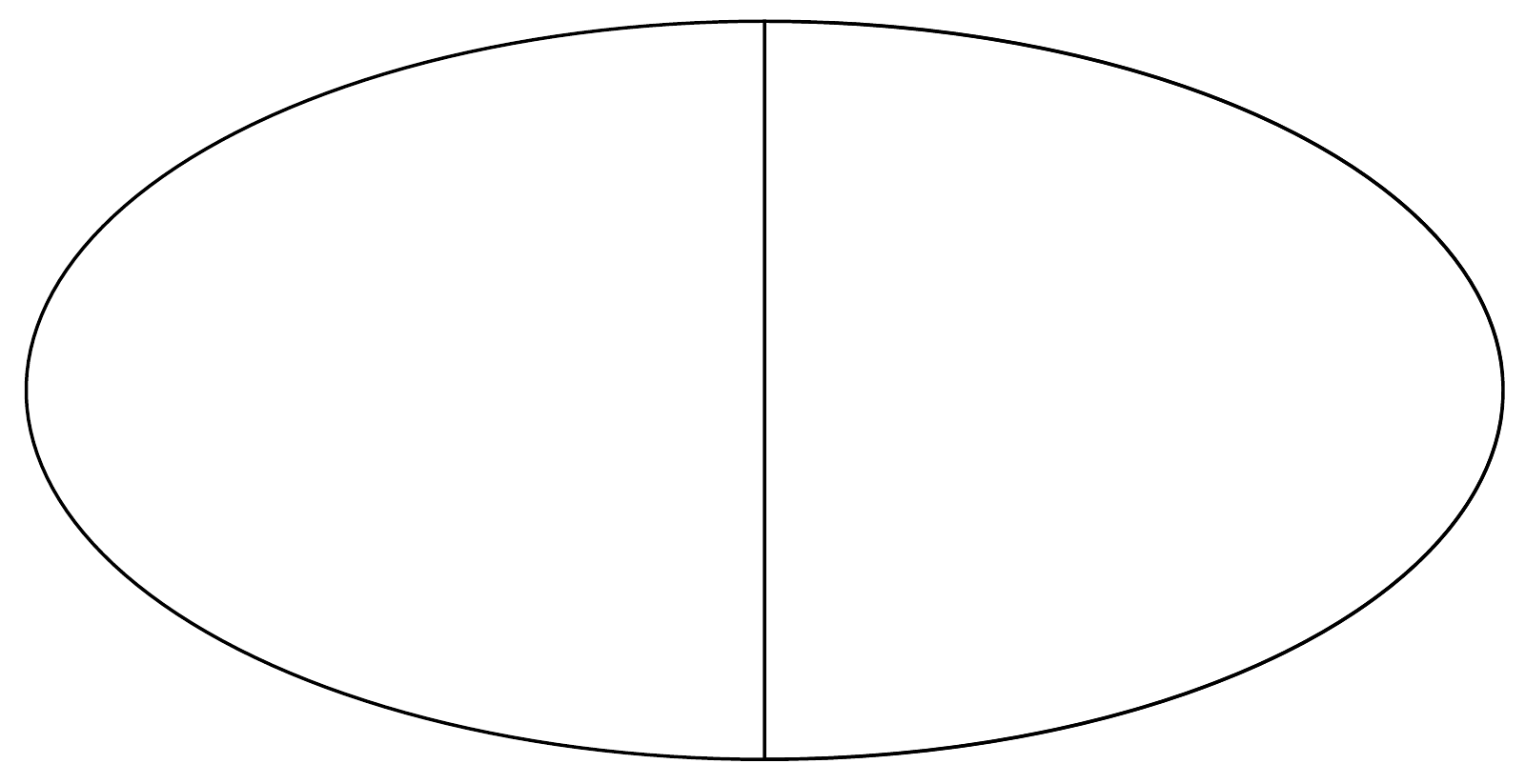}
\caption{Nodal lines separating surplus and deficit regions of sky,
for (left) $\ell=0,\ m=0$ monopole, and $\ell=1,\ m=0$ (middle) and $m=1$ (right) dipoles.}
\label{fig1}
\end{figure}

\begin{figure}[t]
\centering
\includegraphics[width=0.2\textwidth]{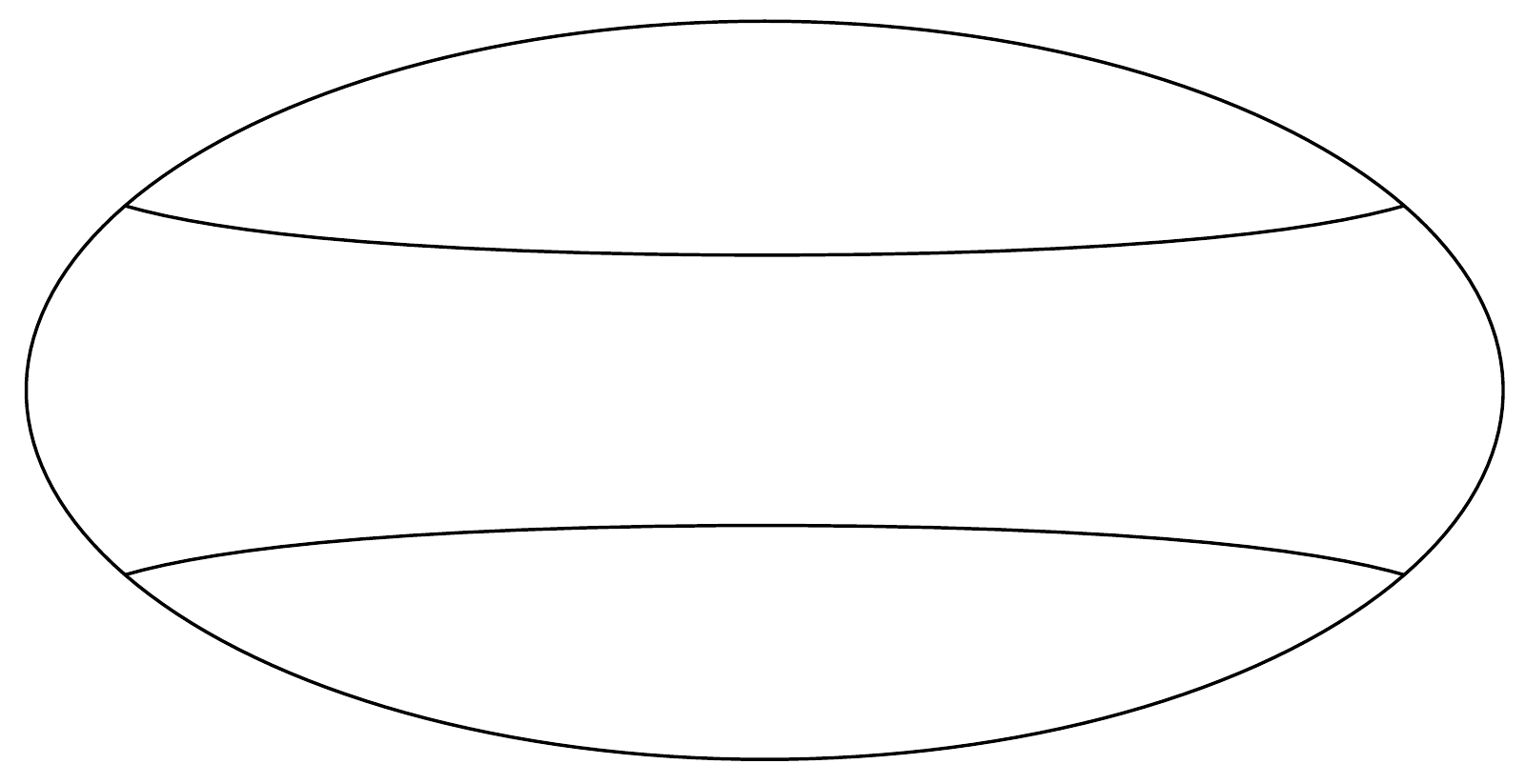}
\includegraphics[width=0.2\textwidth]{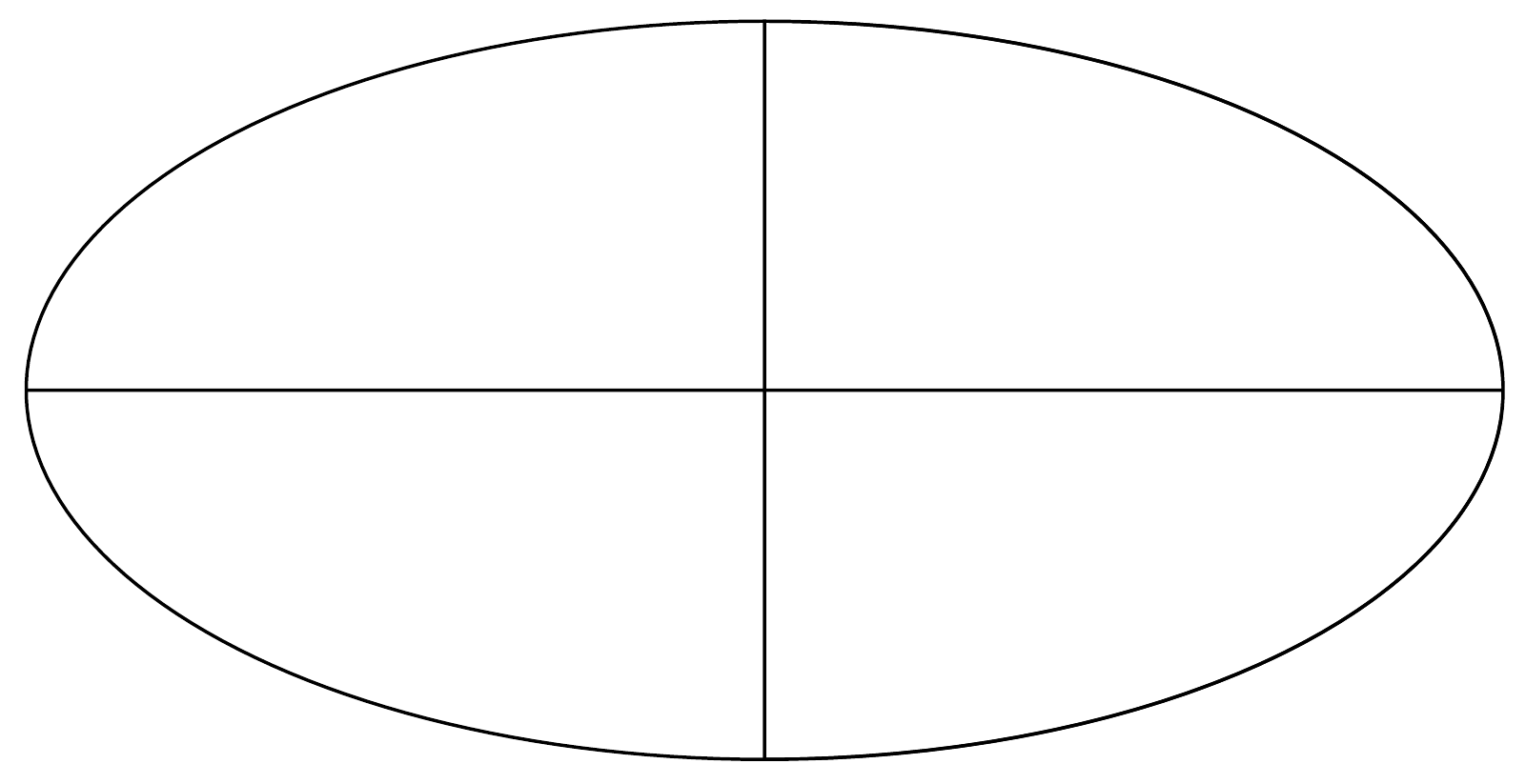}
\includegraphics[width=0.2\textwidth]{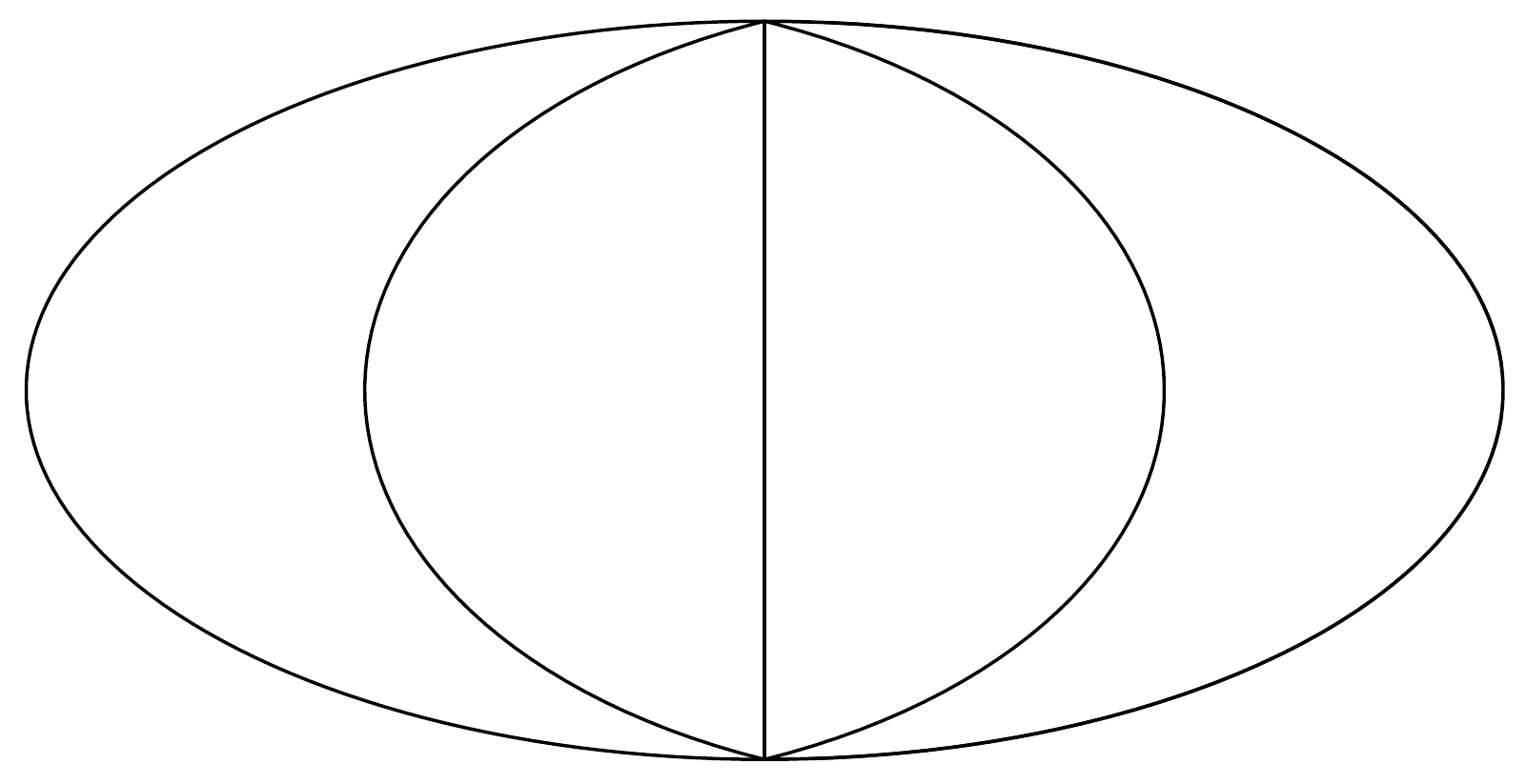}
\caption{Nodal lines separating surplus/deficit regions of sky, 
for $\ell=2,\ m=0,\ 1,\ 2$ quadrupoles, respectively. }
\label{fig2}
\end{figure}

\begin{figure}[t]
\centering
\includegraphics[width=0.2\textwidth]{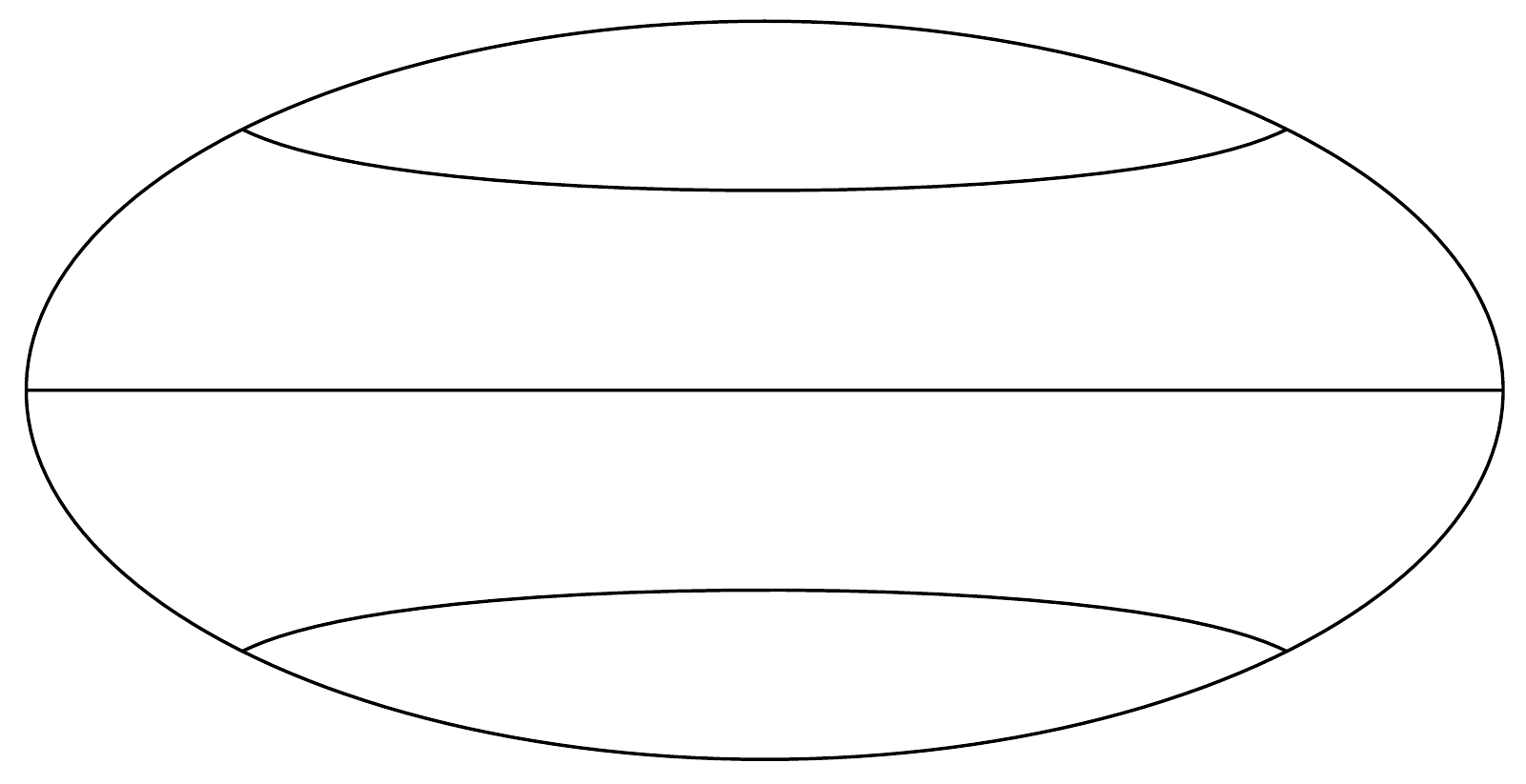}
\includegraphics[width=0.2\textwidth]{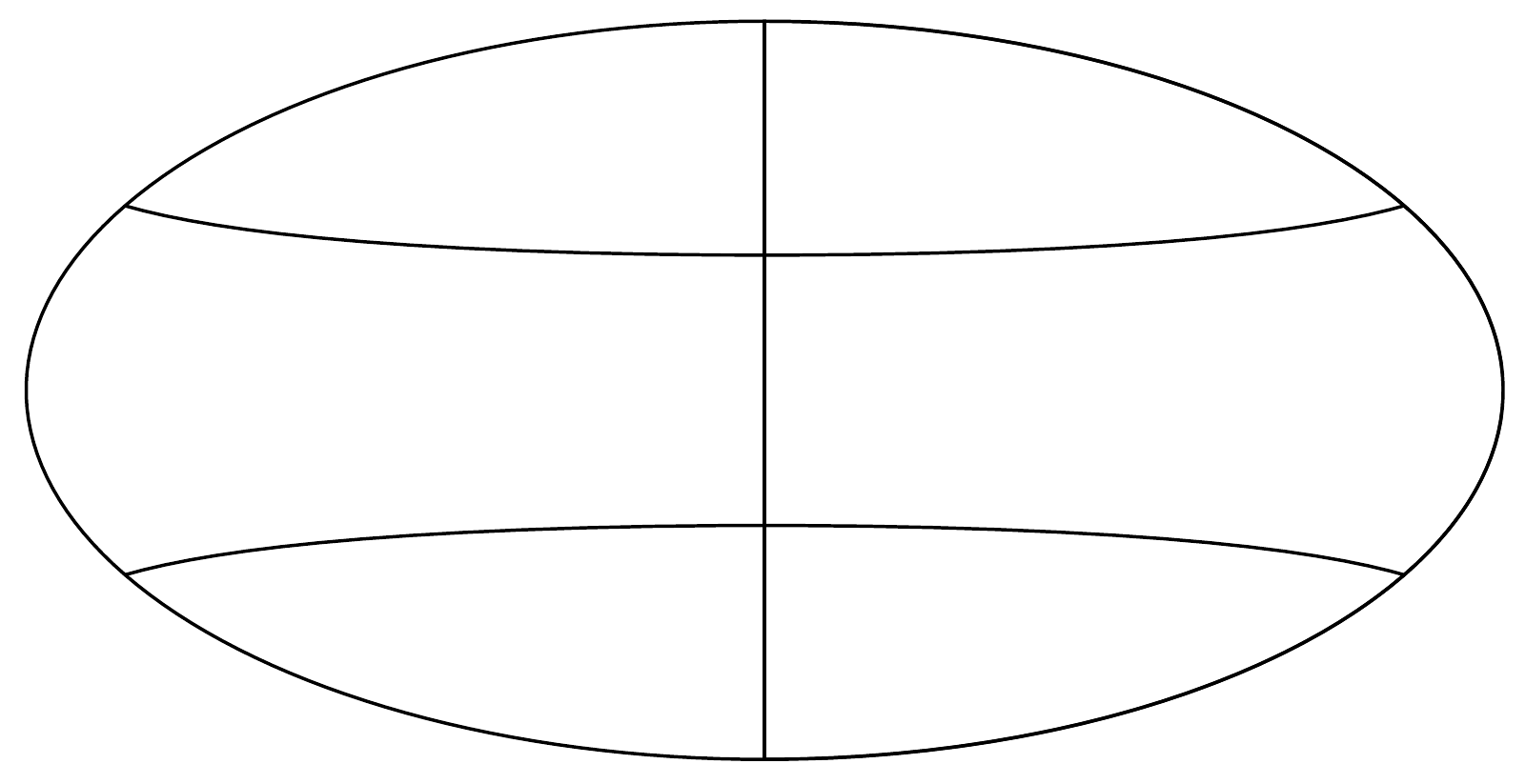} 
\includegraphics[width=0.2\textwidth]{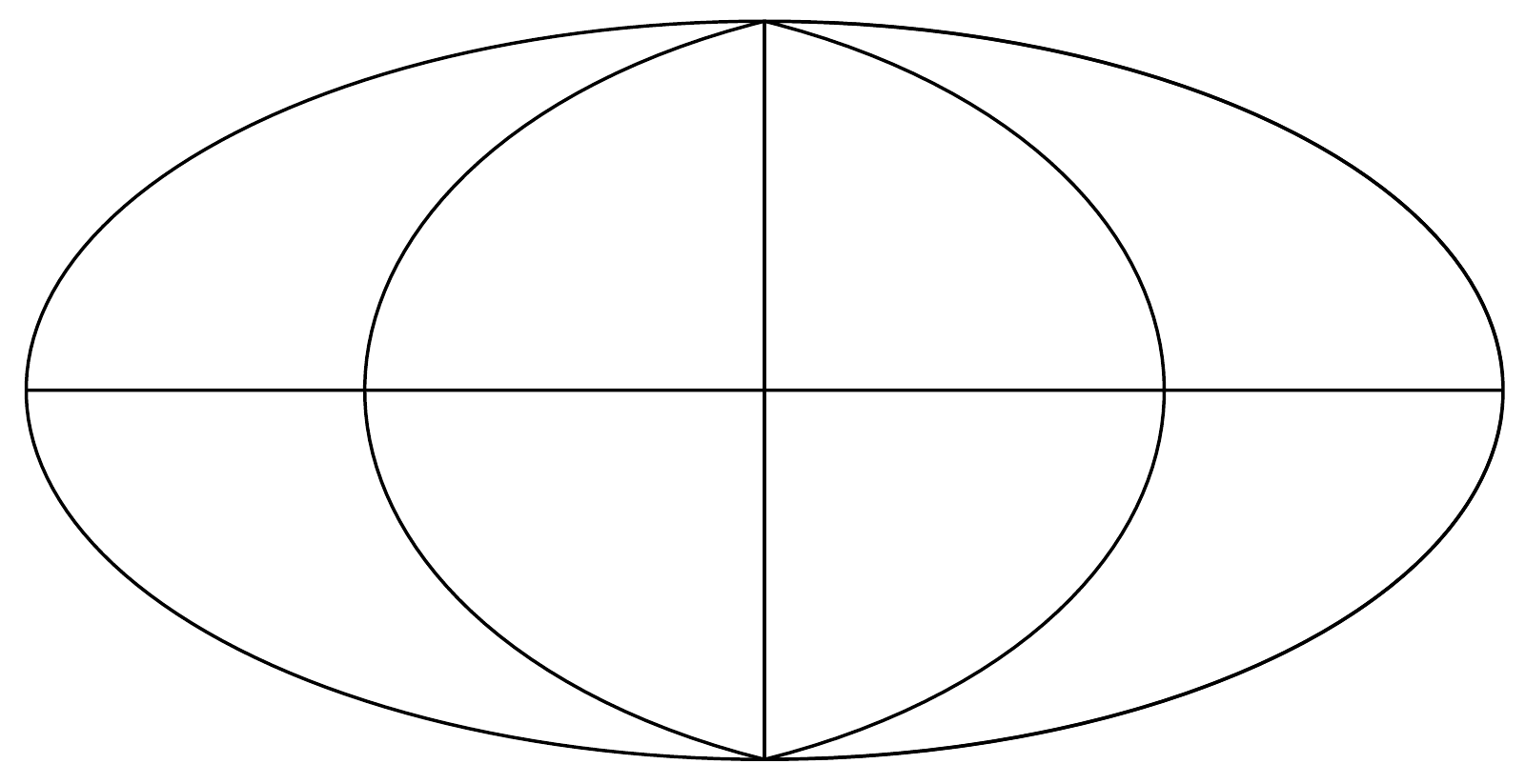}
\includegraphics[width=0.2\textwidth]{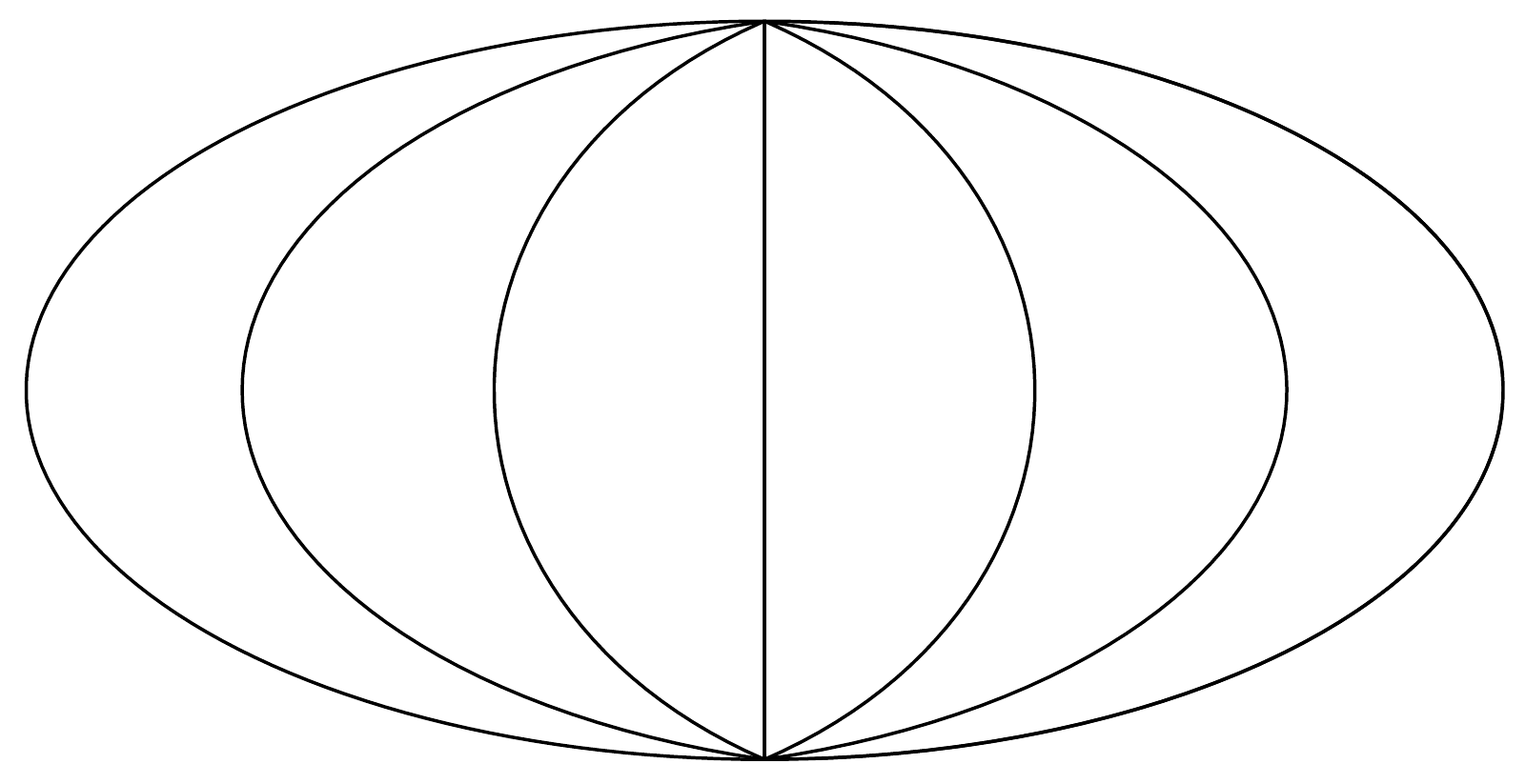}
\caption{Nodal lines separating surplus/deficit regions of sky, 
for $\ell=3,\ m=0,\ 1,\ 2,\ 3$, respectively.}
\label{fig3}
\end{figure}

A nonzero $m$ corresponds to $2\,|m|$ longitudinal ``slices'' ($|m|$ nodal meridians).
There are $\ell+1-|m|$ latitudinal ``zones'' ($\ell-|m|$ nodal latitudes).
In Figs.~(\ref{fig1}-\ref{fig3})
we show the partitioning described by some low-multipole moments.
The configurations with $(\ell,-|m|)$ are related to those with $(\ell,+|m|)$ by a longitudinal phase advance
$\phi\rarr\phi+\frac{\pi}{2}$, or $\cos\phi\rarr\sin\phi$.

\section{Previous anisotropy searches}
The first full-sky large anisotropy search was  based on the combined northern and southern hemisphere
data from the SUGAR and AGASA experiments
taken during a 10~yr period.
Nearly uniform exposure to the entire sky resulted.  
No significant deviation from isotropy was seen, even at energies beyond 
$4 \times 10^{19}$~eV~\cite{Anchordoqui:2003bx}. 
More recently, the Pierre Auger Collaboration carried out various searches for large scale anisotropies in the distribution of arrival directions of cosmic rays above $10^{18}$ eV~\cite{Abreu:2011ve,Auger:2012an}. 

The latest Auger study was performed as a function of both declination and right ascension in several energy ranges 
above $10^{18}$ eV, and reported in terms of dipole and quadrupole amplitudes.
Again no significant deviation from isotropy was revealed. 
Assuming that any cosmic ray anisotropy is dominated by dipole and quadrupole moments in this energy range, the Pierre Auger
Collaboration derived upper limits on their amplitudes. 
Such upper limits challenge an origin of cosmic rays above $10^{18}$ eV from non-transient galactic sources densely distributed in the
galactic disk~\cite{Abreu:2012ybu}.  
At the energies exceeding $6 \times 10^{19}$~eV, however, hints for a dipole anisotropy 
may be emerging~\cite{Anchordoqui:2011ks}.

It must be emphasized that because previous data were so sparse at energies which will be accessible to \J, 
upper limits on anisotropy were necessarily restricted to energies below the threshold of \J.
\J\ expects many more events at $\sim 10^{20}$~eV, allowing an enhanced anisotropy reach.
In addition, \J\ events will have a higher rigidity ${\cal R}=E/Z$, and so will be less bent by magnetic fields;
this may be helpful in identifying particular sources on the sky.

\begin{figure*}[t]
\centering
\includegraphics[width=0.52\textwidth]{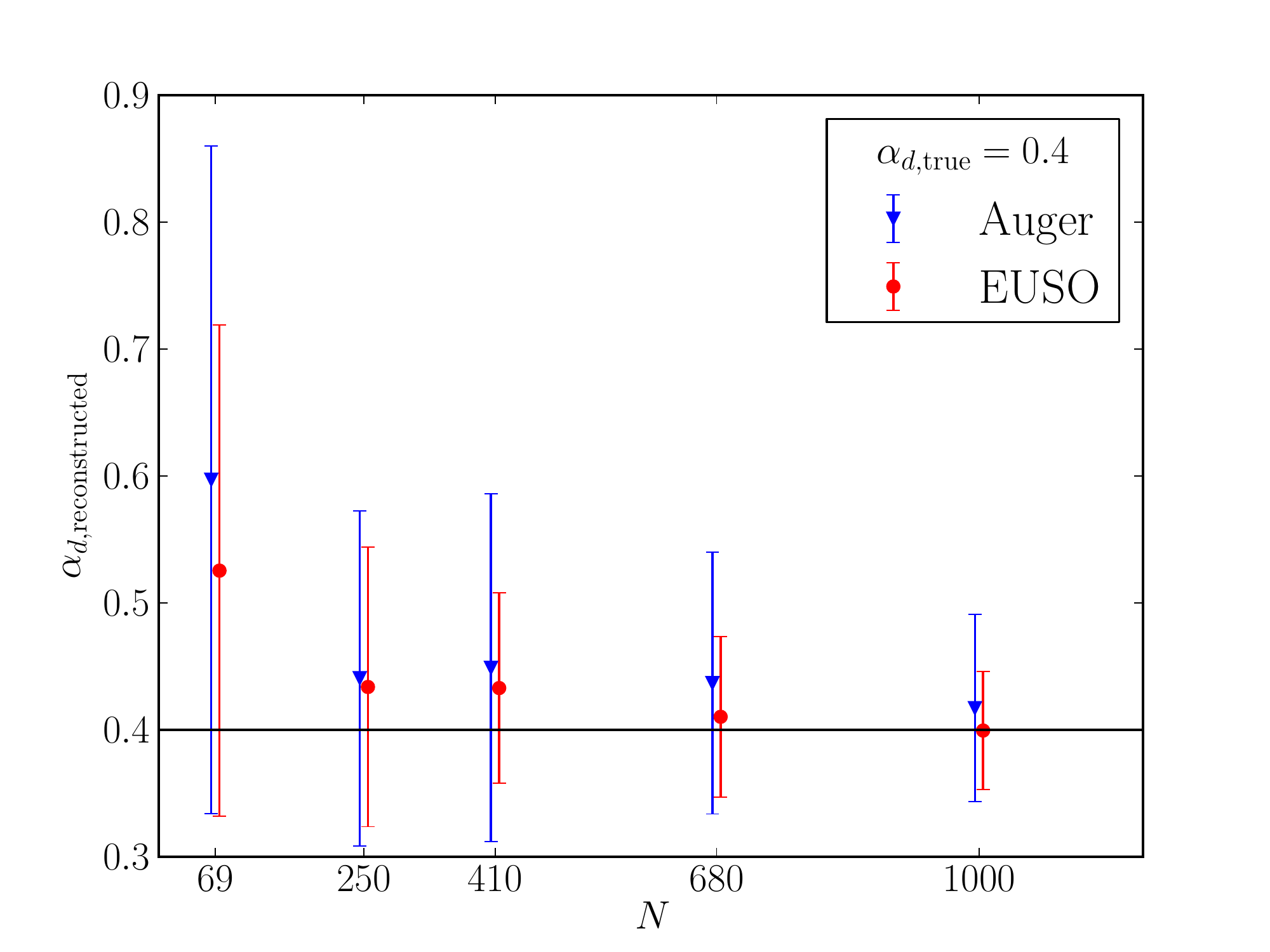}
\includegraphics[width=0.45\textwidth]{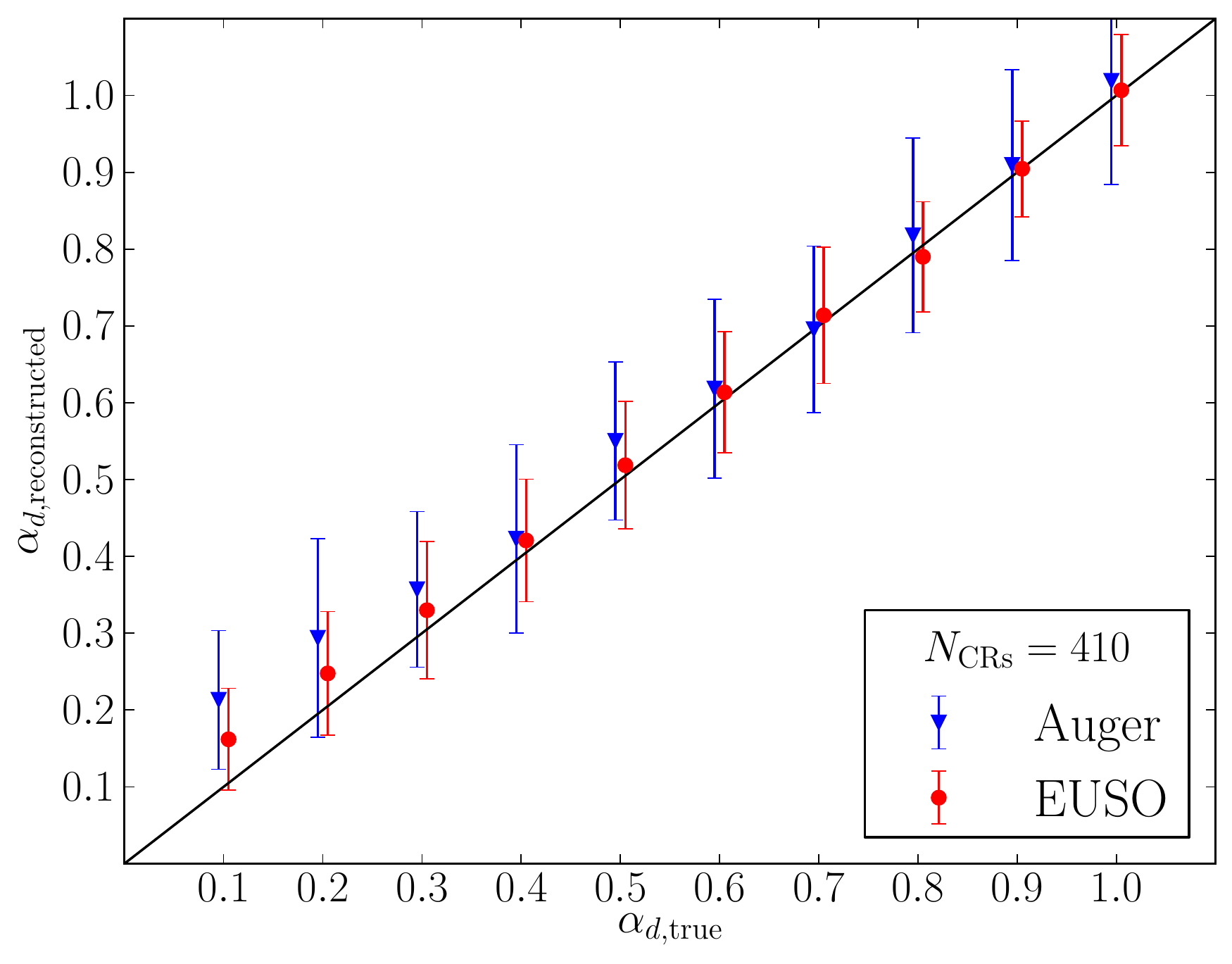}
\includegraphics[width=0.45\textwidth]{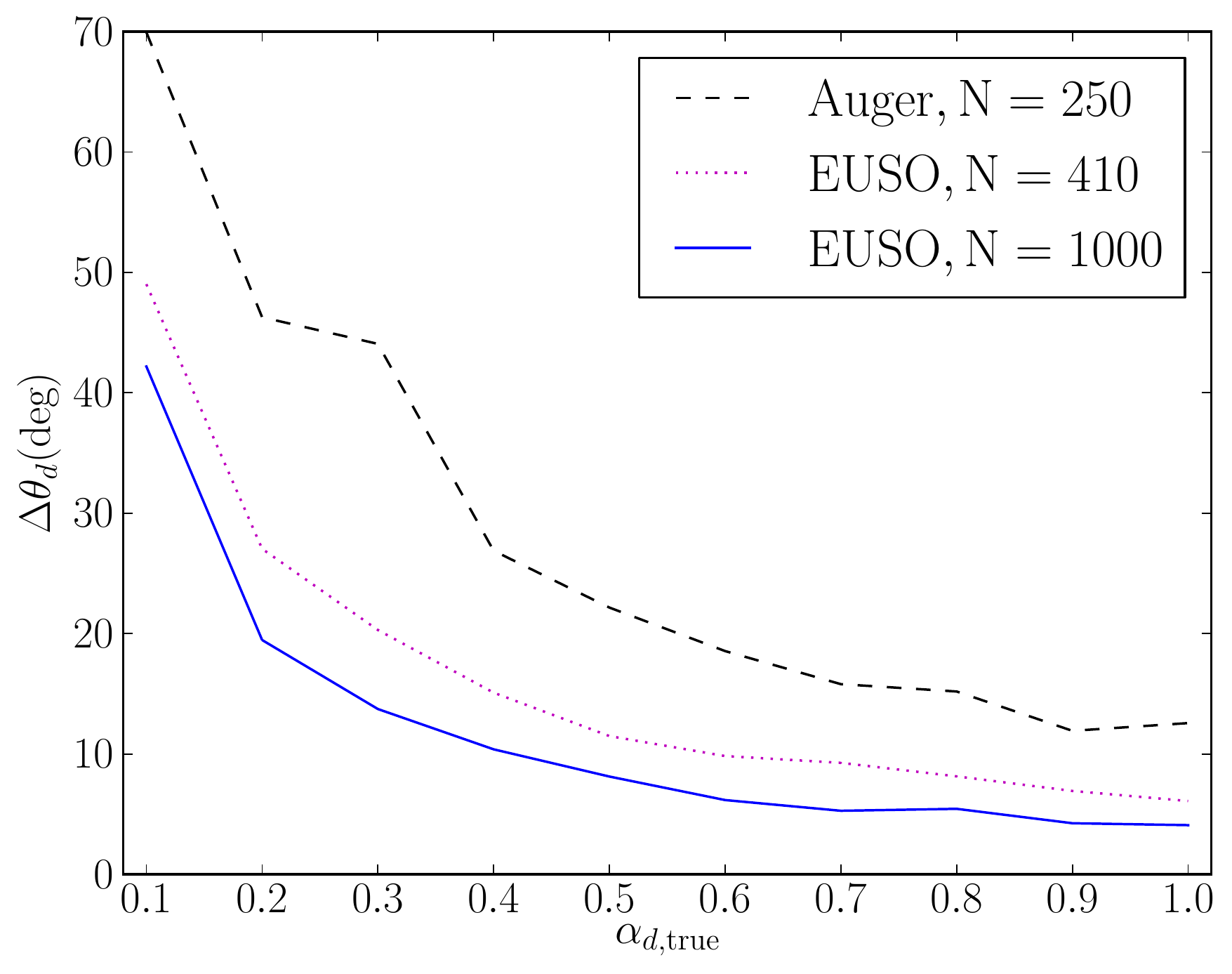}
\includegraphics[width=0.5\textwidth]{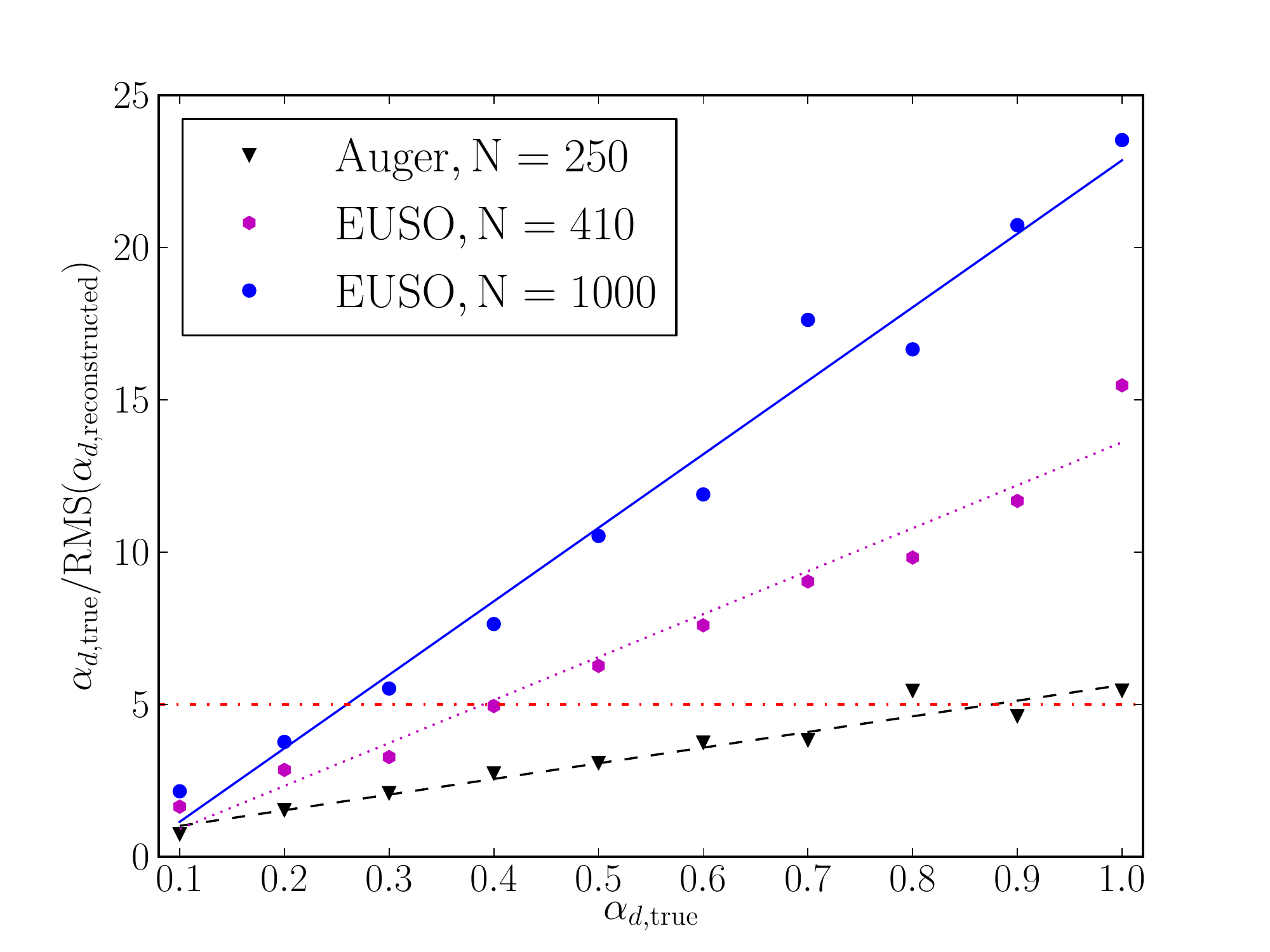}
\caption{Reconstruction of  the dipole amplitude (top panels) and angle (bottom-left panel), for \A\ and \J.
Discovery reach (bottom-right panel) of \J\ and \A, with 5-$\sigma$ horizontal line.} 
\label{fig4}
\end{figure*}
 
\begin{figure*}[t]
\centering
\includegraphics[width=0.5\textwidth]{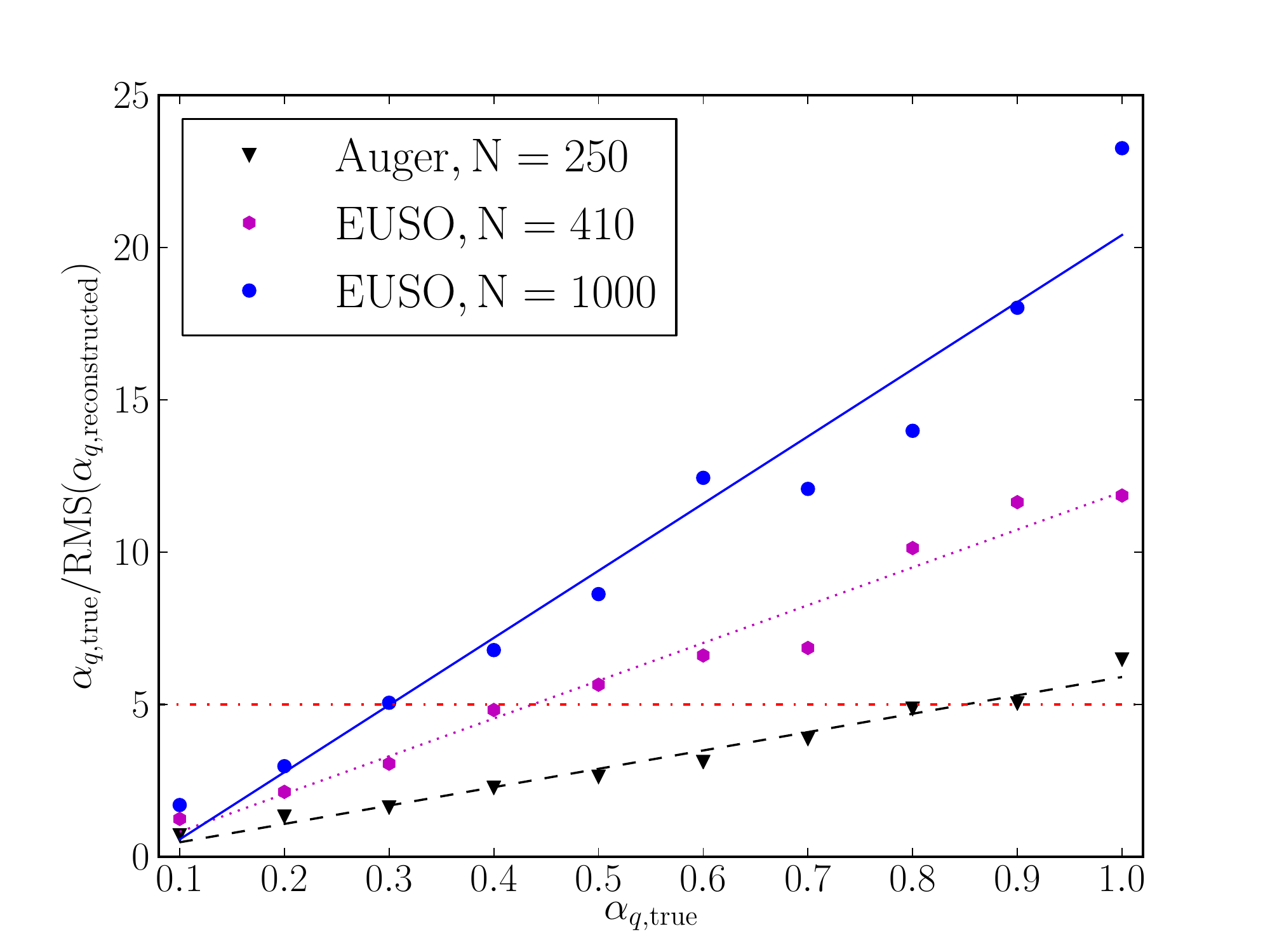}
\caption{Discovery reach of \J\ and \A\ with a 5-$\sigma$ horizontal line.}
\label{fig5}
\end{figure*}

\section{Comparison of all-sky \J\ to half-sky \A}
The ground-based Pierre Auger Observatory (\A) is an excellent, ground-breaking experiment.
However, in the natural progression of science, 
\A\ will be superseded by space-based observatories.
\J\ is designed to be first of its class, building on the successes of \A.

The two main advantages of \J\ over \A\ 
are the (i) greater FOV leading to a greater exposure at EE,
and the (ii) all-sky nature of the orbiting, space-based observatory.  
We briefly explore the advantage of the enhanced exposure first.
We consider data samples of 69, 250, 410, 680, and 1000 events.
The 69 events sample is that presently published by \A\ for events accumulated over three years at and above 55 \EeV.
The annual rate of such events at \A\ is $\sim 69/3=23$.
Thus, the 250 event sample is what \A\ could attain in ten years of running.

Including the \J\ efficiency down to 55 \EeV\ reduces the factor of 9 relative to \A\ down
to a factor $\sim$6 at and above 55 \EeV. 
We arrive at the 410 event sample as the \J\ expectation at and above 55 \EeV\ after three years running in nadir mode
(or, as is under discussion, in tilt mode with an increased aperture but reduced PDM count).
A 680 event sample is then expected for five years of \J\ in a combination of nadir and tilt mode.
Finally, the event rate at an energy measured by High Resolution Fly's Eye (HiRes) is known to exceed that of \A\ by~50\%.
This leads to a five-year event rate at \J\ of about 1000 events.  

Now we turn to the $4\pi$~advantage.
Commonly, a major component of the anisotropy is defined via a max/min directional asymmetry,
$\alpha\equiv \frac{I_{\max} - I_{\min}}{I_{\max} + I_{\min}}$.
For a monopole plus dipole distribution $1+\alpha_D\cos\theta$, one
readily finds that $\alpha=\alpha_D$.
For a monopole plus quadrupole distribution  $1-C\cos^2\theta$ (no dipole),
one finds 
that  $\alpha=\frac{C}{2-C}$, and 
$C=\frac{2\alpha}{1+\alpha}$.

In Fig.~(\ref{fig4}) we compare the capability of \J\ and \A\ to reconstruct a dipole anisotropy.
In this comparison, both advantages of \J,
namely the increased FOV and the $4\pi$~sky coverage, are evident.

Dipole (plus monopole) plots are composed in the following way.
First, we choose a dipole amplitude $\alpha_{\rm true}$ (relative to the monopole amplitude),  and a dipole direction.
The latter is randomly oriented, and defines the axis for the polar angle $\theta$. 
Then, a given number of events, 250, 410, or 1000, are randomly distributed within the weighting
$1+\alpha_{\rm true}\cos\theta$.
Next, the fitting algorithm determines, as best it can, reconstructed values for $\alpha$ and for the dipole direction.
This process is repeated 100 times, each time with a different randomly oriented dipole direction.
Results are averaged, and presented in the figures.
This formulation results in our observation point, the Earth, being located at the center of the dipole distribution.
A dipole distribution might be indicative of a single, dominant cosmic-ray source in the presence of a magnetic field.

In the top-left panel of Fig.~(\ref{fig4}) are shown the error bars that result from a reconstruction of the dipole amplitude
(we have chosen $\alpha_{\rm true}=0.4$ for illustration), for the various $N$-event samples.
The errors in $\Delta\alpha$ are seen to scale as $1/\sqrt{N}$
as one would expect for a Poisson distribution of events in $\cos\theta$.
More significantly, the reconstruction errors in \A\ for the dipole amplitude are almost twice those of \J, 
due to the limited sky-coverage of \A\ (and even worse for the quadrupole, to be analyzed next)~\cite{ParizotMethod}.  
Moreover, if the dipole were aligned with the zenith angle of \A, 
then \A\ could easily confuse a quadrupole with a half-dipole.
There is no such ambiguity with the $4\pi$ coverage of \J.

The top-right panel of Fig.~(\ref{fig4}) shows the error in reconstruction of the dipole amplitude,
for a fixed event number $N=410$~events.  
Of course, \A\ will not attain an event sample of 410, but the figure correctly displays the 
loss of quality when the acceptance is reduced from the all-sky $4\pi$ steradians to that of \A,
even for the same number of events as \J.
The bottom-left panel of Fig.~(\ref{fig4}) shows the reconstruction errors on the dipole direction,
versus the dipole amplitude, for three event samples, \A\ with 250 events,
and \J\ with 410 and 1000 events.
Not surprisingly, the ability to reconstruct the dipole direction improves dramatically with
an increase in the dipole amplitude.
For $\alpha$ exceeding 0.2, the 410 event \J\ sample has less that half the error of the 
250 event \A\ sample,
and the 1000 event sample has less than a third the error of the \A\ sample.

The final panel in Fig.~(\ref{fig4}) gives the number of $\sigma$ for reconstruction of the dipole amplitude,
as a function of the true dipole amplitude.
The discriminatory power of $4\pi$ \J\ is obvious.
A discovery claim (5-$\sigma$) is evident to \A\ only for dipole amplitudes above 0.80.
However, 410-event \J\ (3yrs) can claim discovery for an amplitude down to 0.40,
and 1000-event \J\ can claim discovery all the way down to 0.28.
The latter sample can reveal 3-$\sigma$ ``evidence'' for an 
nonzero dipole amplitude down to 0.20.

\begin{figure*}[t]
\centering
\includegraphics[height=0.12\textheight,width=0.38\textwidth]{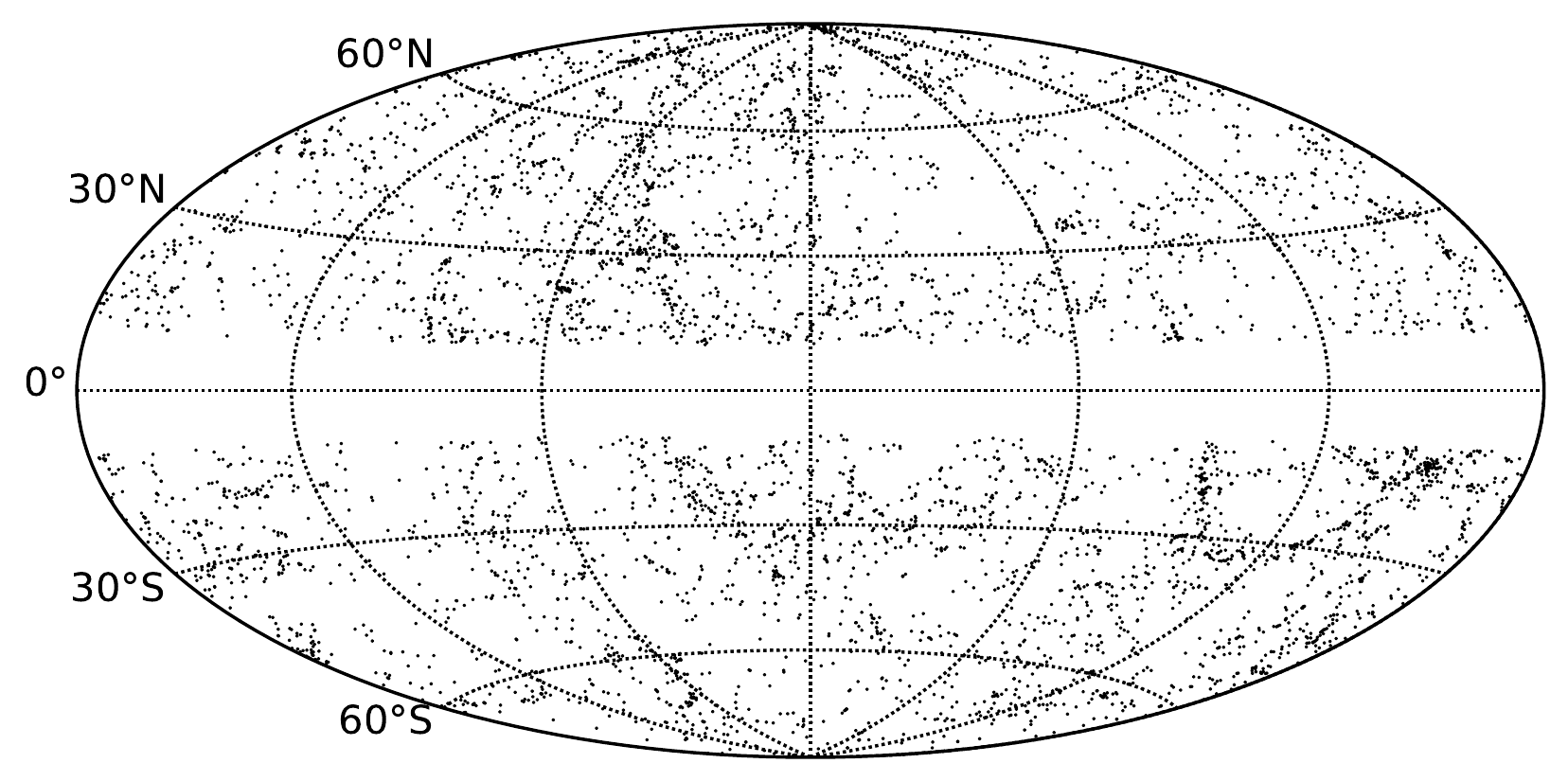}
\includegraphics[height=0.12\textheight,width=0.28\textwidth]{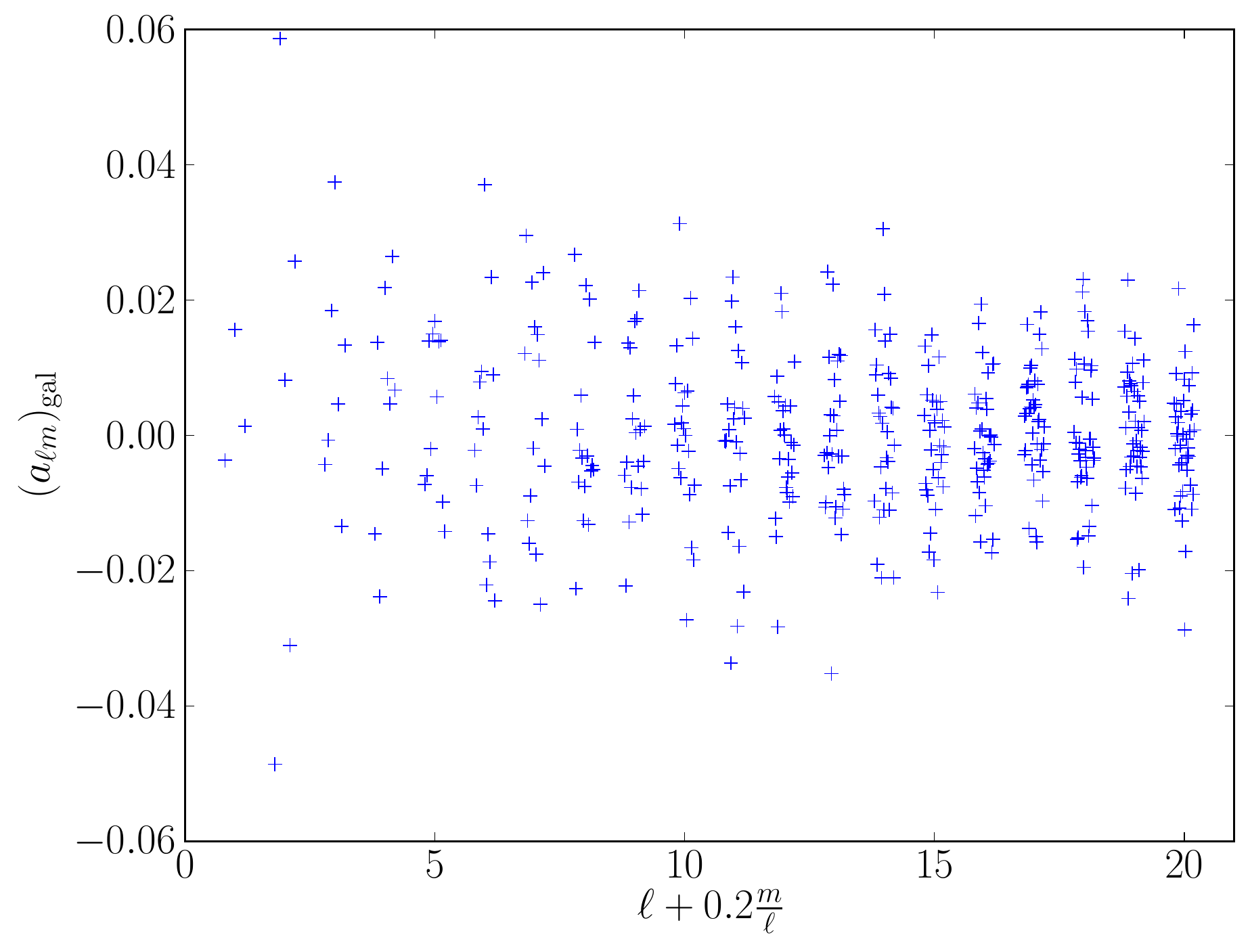}
\includegraphics[height=0.12\textheight,width=0.28\textwidth]{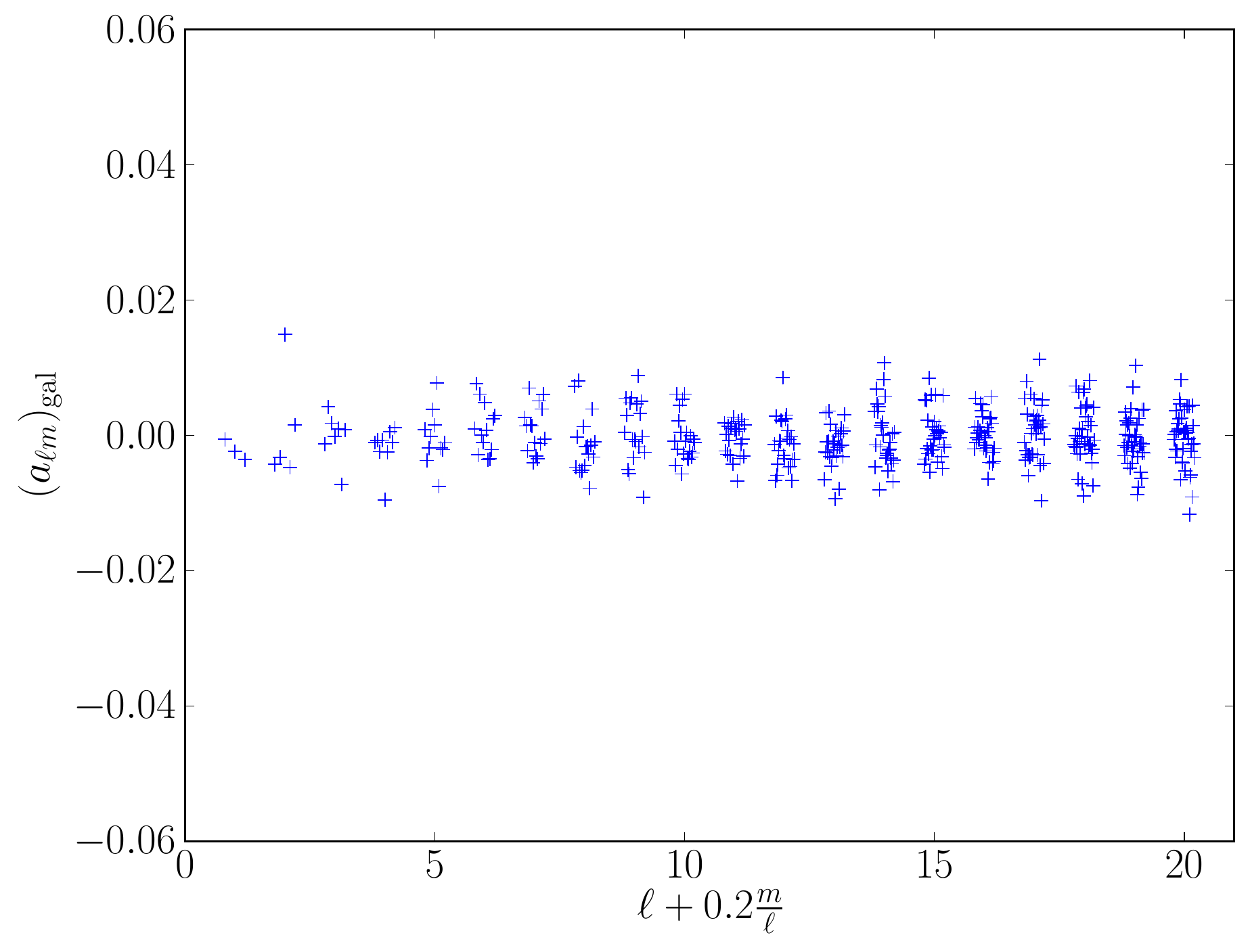}
\caption{
Sky-map of (left panel) the ``5310 Galaxy-source'' distribution, taken from the 2MRS
catalog of galaxies~\cite{Huchra:2011ii} out to z=0.03 (115~Mpc),
excluding the Galactic plane at $|b| \le 10^\circ$; and the fitted $a_{\ell m}$ values (center panel).
(right panel) $a_{\ell m}$'s from randomly distributed 5310-event ``isotropic'' data,
excluding the same strip of Galactic plane. The $\ell/m$-dependency of the abscissa in
the $a_{\ell m}$ plots is chosen~\cite{Sommers:2000us} to increase the visibility
of the $(2\ell+1)\ m$ values at fixed $\ell$.
The $|b|\le10^\circ$ cut is accounted for analytically in each of the right two figures.
Note the large $|a_{\ell=2}|$ values in the center figure.
}
\label{fig6}
\end{figure*}

In Fig.~(\ref{fig5}) we repeat the reconstruction comparison of \J\ and \A,
but this time for a quadrupole distribution.
Our results are an average over 100 trials of a random but weighted distribution of a 
purely isotropic monopole and anisotropic quadrupole, the latter having a randomly chosen orientation,
again with the Earth located at the center of the distribution.
There is no dipole in the input data set.
A quadrupole amplitude might be indicative of a dominant Galactic distribution of sources,
or even of a distribution of dominant sources in the Supergalactic Plane.
A general quadrupole ($\ell =2$) has five allowed $m$ values.
Here we consider a pure $m=0$ distribution of events, 
i.e., a quadrupole with azimuthal symmetry about the quadrupole axis, 
as shown in the first sky-map of Fig.~(\ref{fig2}).  
We have again chosen $\alpha=0.4$ for illustration.
The total distribution of events mimics an oblate spheroid with relative probability distribution
$1-\frac{4}{7}\cos^2\theta$.

The same analysis as in Fig.~(\ref{fig4}) was done for the quadrupole case.
A partial sky observatory fits a quadrupole much more poorly. 
Fig.~(\ref{fig5}) shows that \J\ can claim 5-$\sigma$ discovery of a quadrupole amplitude 
as low as 0.4 with 410 events, and as low as 0.3 with 1000 events.
Also with 1000 events, \J\ is sensitive to a 3-$\sigma$ indication down to an amplitude of 0.2.
On the other hand,  \A\ is incapable of claiming a quadrupole discovery 
unless the quadrupole amplitude is maximum, an unlikely value.

\section{Future studies}

In the near future, we will include several additional complicating, real world aspects of the 
spherical harmonic search for anisotropy.
One relates to the energy resolution for individual events.
With a spectrum falling steeply in energy, 
a spill-over of a lower-energy bin with isotropic events into a higher-energy bin 
with potentially anisotropic events will dilute the signal.
While more study of this issue is warranted, this seems not to be a serious concern.
\A\ quotes an energy resolution $\Delta E/E$ of 22\%, with 12\% statistical, and 
about 20\% systematic.  
Simulations in \J\  to determine the energy resolution at extreme energies are ongoing,
with the present upper limit $\sim 30\%$.
Also, the data sample of \J\ used in anisotropy studies is not bound to the 55 \EeV\ threshold that 
\A\ chose due to its limited statistics.  With more statistics expected, \J\ can choose a higher-energy threshold.
Of course, any anisotropy will turn on gradually in energy, and simulations must include this fact.

Other probable non-concerns are the systematic error in the angular resolution of \J\ ($\lesssim 3^\circ$),
and the bending of proton trajectories at extreme energies, 
given firstly by the random walk equation through extragalactic magnetic domains of strength 
$B_{\rm nG}$ in units of nanoGuass, and coherence size $\lambda$,
\beq{magbending}
\delta\theta^\circ = 0.8\,Z\, \left( \frac{B_{\rm nG}}{E_{20}}\right)\, \sqrt{\frac{D\,\lambda}{10\,{\rm Mpc}^2}}\,;
\eeq
where Z and $E_{20}$ are the CR charge and energy in units of 100 \EeV, respectively, 
and $D$ is the distance traveled by the CR.
One sees that for a proton ($Z=1$) at 100 \EeV, the natural unit of bending is a degree
in the extragalactic magnetic field.
On the other hand, heavy nuclei trajectories may be so severely bent as to eliminate 
even large-scale event anisotropies.  
So our hope hangs on protons being dominant at extreme energies.
Complicating the issue is that the bending may be more or less if
filamentary structure or voids are encountered en route.
Subsequent to the transit of extragalactic space, the CR encounters the Galactic magnetic field,
which is larger than than the extragalactic field, but is also known more precisely.

Our near-future studies will incorporate 
energy resolution effects,
realistic estimates of Galactic and extragalactic magnetic fields, and ``GZK'' energy
losses on cosmic radiation fields.
To incorporate these effects, we will add energy and direction assignments
for the individual simulated events in the context of two models, which we call 
the ``Galaxy source model'' and the ``Single source model'', the latter being motivated by weak evidence
for a UHECR excess in the direction of Centaurus A.
For the Galaxy source model, the simulated source data 
is weighted to the 2MRS all-sky catalog of galaxies out to $z=0.03$ (about 115~Mpc)~\cite{Huchra:2011ii}.  
For the single source model, the simulated source data is fixed to a single source on the sky.
For each data set, we will propagate the CRs to Earth and  
perform a multipole analysis of the resulting sky map.

The flavor of our work in progress can be gleaned from Fig.~(\ref{fig6}).
The first panel in Fig.~(\ref{fig6}) presents the sky-map of the 5310 galaxies 
present in the 2MRS survey, which reaches out to z=0.03, corresponding to a distance
$\sim z/H_0\sim115$~Mpc.
In the middle panel 
are shown the $a_{\ell m}$'s that result from the Galaxy-source model.
As a control, the right panel displays the $a_{\ell m}$'s that result from an isotropic
distribution of 5310 events with the same $y$ scale.
In both analyses, the Galactic Plane at declination below $|b| \le 10^\circ$ is omitted.
An eyeball comparison of the two panels shows the power of the $a_{\ell m}$'s to reveal anisotropy.
In particular we recover the predicted quadrupole structure in that the largest
$|a_{\ell m}|$ are in the $\ell=2$ column.

\section{Conclusions}
The two main advantages of space-based observation of EECRs over ground-based observatories 
are increased FOV and $4\pi$ sky coverage with uniform systematics.
The former guarantees increased statistics,
whereas the latter enables a partitioning of the sky into spherical harmonics.  
We have begun an investigation, using the spherical harmonic technique, 
of the reach of \J\  into potential anisotropies in the EECR sky-map.  
The discovery of anisotropies would help to identify the long-sought source(s) of EECRs.

\ack
This work is supported in part by a Vanderbilt Discovery Grant (TJW and PBD), 
NSF CAREER PHY-1053663 and NASA 11-APRA11-0058 Awards (LAA),
Alfred P. Sloan Foundation (AAB), and NSF GAANN fellowship (MR).

\section*{References}
\bibliographystyle{iopart-num}
\bibliography{CR}

\end{document}